\documentclass[reprint,
 amsmath,amssymb,
 aps,
]
{revtex4-1}

\usepackage{graphicx}
\usepackage{dcolumn}
\usepackage{bm}
\usepackage{subfigure}

\begin{document}


\title{Theoretical calculation of the fine-structure constant\\ and the permittivity of the vacuum}

\author{G. B. Mainland}%
 \email{mainland.1@osu.edu}
\affiliation{Department of Physics, The Ohio State University at Newark, 1179 University Dr., Newark, OH 43055}

\author{Bernard Mulligan}%
 \email{mulligan.3@osu.edu}
\affiliation{Department of Physics, The Ohio State University, Columbus, OH 43210}


\author{}
\affiliation{}

\setcounter{equation}{0}

\date{\today}

\begin{abstract}
Light traveling through the vacuum interacts with vacuum fluctuations  similarly to the way that light traveling through a dielectric interacts with ordinary matter.  And just as the permittivity of a dielectric can be calculated, the permittivity $\epsilon_0$ of the vacuum can be calculated, yielding an equation for the fine-structure constant $\alpha$. The most important contributions to the value of  $\alpha$ arise from the interaction of photons with charged lepton-antilepton vacuum fluctuations that appear in  the vacuum as on-shell, bound states.  Considering these contributions only to first order in alpha, the  fully screened $\alpha \cong 1/(8^2\sqrt{3\pi/2}) \cong 1/139$.  
\end{abstract}


\maketitle 


The fine-structure constant $\alpha$, which has a value independent of the choice of units, is given by \footnote{SI units are used throughout.}
\begin{equation}\label{eqn:1}
\alpha = \frac {e^2}{(4 \pi \epsilon_0)\hbar c}= \frac {e^2}{2 \epsilon_0h c}=\frac{e^2}{2h}\sqrt{\frac{\mu_0}{\epsilon_0}}\,,
\end{equation}
where $c=1/\sqrt{\mu_0\epsilon_0}$ is used to obtain the final equality. In \eqref{eqn:1} $e,  h, c, \epsilon_0$, and $\mu_0$ are, respectively, the (screened) magnitude of the charge on an electron, Planck's constant, the speed of light in the vacuum, the permittivity of the vacuum, and the magnetic constant.                                                                                                  

Using  techniques  similar to those employed for calculating the permittivity of a dielectric,  a formula for the permittivity $\epsilon_0$  of the vacuum is derived. The calculation  is simplified\textendash and the numerical accuracy is reduced\textendash by including only the most significant interactions, those of photons interacting with charged lepton-antilepton vacuum fluctuations.  The formula for $\epsilon_0$ is easily converted into a formula for $\alpha$, yielding the approximate theoretical value $1/\alpha \cong  8^2\sqrt{3\pi/2)}  \cong  139$, which is to be compared with the experimental value $1/\alpha=1/137.036\dots$. Solving for the permittivity of the vacuum,  $\epsilon_0 \cong (4^4)\sqrt{6\pi}\,e^2/hc \cong 8.98\times 10^{-12} {\rm C^2}/{\rm Nm^2}$. The experimental value is $8.85\times 10^{-12} {\rm C^2}/{\rm Nm^2}$.

The possibility that the properties of the quantum vacuum determine, in the vacuum, the speed of light and the permittivity\cite{Leuchs:10,Leuchs:13,Urban:13} date back to the beginning of quantum mechanics. As early as 1936 the idea of treating the vacuum as a medium with electric and magnetic polarizability was discussed by Weisskopf and Pauli \cite{Weisskopf:94, Pauli:94}.   Twenty-one years later Dicke  \cite{Dicke:57} wrote about the possibility that the vacuum could be considered as a dielectric medium.   Einstein's 1920 Leiden lecture ``Ether and the Theory of Relativity'' \cite{Einstein:20} was a significant influence on Wilczek's  2008 book {\it The Lightness of Being} \cite{Wilczek:08} that met Einstein's challenge by expressing and encompassing the fundamental characteristics of space and time through the concept of the Grid, ``the entity we perceive as empty space.  Our deepest physical theories reveal it to be highly structured; indeed, it appears as the primary ingredient of reality.'' From Wilczek's conceptualization it is expected that the Grid determines all physics. In this paper a derivation of the value of  $\alpha$ is carried out based on characteristics of the Grid.  A second foundation of this calculation is the Heisenberg uncertainty relation, $\Delta E \Delta t \geq\hbar/2$, where $\Delta E$ and $\Delta t$ are the respective,  simultaneous uncertainties in the energy and time.  

Here calculation of the permittivity $\epsilon_0$ of the Grid\textendash and of $\alpha$\textendash is begun  by considering the classical  formula \cite{Kramers:24,Rossi:59, Feynman:63}  for the  permittivity $\epsilon$ of a dielectric.  Vacuum fluctuations, while they exist,  are real, physical particles

To determine how the formula must be modified to describe oscillating vacuum fluctuations instead of physical particles that oscillate, the derivation of the classical formula \cite{Rossi:59} is briefly reviewed. 

Consider a plane, sinusoidal, linearly-polarized electromagnetic wave traveling in the z-direction with the electric field along the x-axis, the magnetic field along the y-axis, and both fields oscillating at an angular frequency $\omega$.  The origin of the reference frame is at the equilibrium position of the dipole charges. The  $j^{\rm th}$ variety of oscillators  is composed of charges $q_j$ and $-q_j$ that oscillate along the x-axis, affecting the progress of the light wave. Writing $F_x=ma_x$ the following equation is obtained\cite{Rossi:59}
\begin{equation}\label{eqn:2}
-kx(t)-(h_r+h_c)\frac{{\rm d}x(t)}{{\rm d}t}+q_jE\cos\omega t =\mu_j \frac{{\rm d}^2x(t)}{{\rm d}t^2}\,.
\end{equation}
On the left-hand side of the above equation $-kx(t)$ is the elastic restoring force, $-(h_r+h_c)\frac{{\rm d}x(t)}{{\rm d}t}$ is the damping force resulting from radiation ($h_r$)  and collisions ($h_c$),  $q_jE\cos\omega t$ is the force on the $j^{\rm th}$ oscillator resulting from the electric field $E\cos\omega t$ of the light wave, and $\mu_j$ is the reduced mass of the $j^{\rm th}$ oscillator.  Defining the electric dipole moment $p_j(t)$, the damping parameter $\tau$ and the classical resonant frequency $\omega_0^j$ of the oscillator, respectively, by
\begin{equation}\label{eqn:3}
p_j(t)=q_jx(t) ,\; \frac{1}{\tau}= \frac{h_r+h_c}{\mu_j}\;\,{\rm and}\;\,(\omega_0^j)^2=\frac{k}{\mu_j}\,,
\end{equation}
\eqref{eqn:2}  can be rewritten as
\begin{equation}\label{eqn:4}
 \frac{{\rm d}^2p_j(t)}{{\rm d}t^2} +\frac{1}{\tau}\frac{{\rm d}p_j(t)}{{\rm d}t}+(\omega_0^j)^2p_j(t)= \frac{q_j^2}{\mu_j}E\cos\omega t \, .
\end{equation}

The solution of  \eqref{eqn:4} is of the form
\begin{equation}\label{eqn:5}
p_j(t)=p_j^{(0)}\cos(\omega t-\phi) \, ,
\end{equation}
where 
\begin{subequations}\label{eqn:6}
\begin{equation}\label{eqn:6a}
p_j^{(0)}=\frac{(q_j^2/\mu_j)E}{\sqrt{[(\omega_0^j)^2-\omega^2]^2+\omega^2/\tau^2}}\,,
\end{equation}
and
\begin{equation}\label{eqn:6b}
\tan \phi=\frac{\omega}{\tau[(\omega_0^j)^2-\omega^2]}\,.
\end{equation}
\end{subequations}
Typically in dielectrics the damping is small so the damping term $\omega^2/\tau^2$ can be neglected. Except when the resonant frequency $\omega_0^j$ and $\omega$ are almost the same, it follows from  \eqref{eqn:6b} that the electric field and the dipole have almost the same phase when $\omega < \omega_0^j$  and essentially opposite phases when $\omega > \omega_0^j$. Therefore,
\begin{equation}\label{eqn:7}
p_j^{(0)} \cong \frac{(q_j^2/\mu_j)E}{(\omega_0^j)^2-\omega^2}\,.
\end{equation}

Using $\epsilon E=\epsilon_0 E +P$, where  the permittivity of the dielectric is $\epsilon$,  the polarization density $P=\Sigma_j N_j p_j^{(0)}$, and  $N_j$ is the number of oscillators per unit volume of the $j^{\rm th}$ variety that are available to interact, it follows that \cite{Jackson:99}
\begin{equation}\label{eqn:8}
\epsilon   \cong   \epsilon_0   \, +  \sum_j \frac {N_j (q_j^2/\mu_j)}{ (\omega_{0}^j)^2 - \omega^2 }  \, .
\end{equation}
The oscillators in the dielectric contribute to an increase in $\epsilon$ from the value $\epsilon_0$ in the Grid. 

The quantum formula \cite{Holstein:92} for the propagation of a photon through a dielectric is identical to \eqref{eqn:8} except  that $\omega_{0}^j$  now is the frequency corresponding to the ground state instead of the classical resonant frequency.  The above formula can be used for gases: complicating issues can arise when calculating the permittivity of liquids or solids.    In \eqref{eqn:8}  the second term on the right-hand side is the increase in the permittivity from  $\epsilon_0$ to $\epsilon$  as a result of photons interacting with oscillators in the dielectric and results entirely from polarization of the atoms, molecules or both in the dielectric. It then follows that the permittivity of the Grid must result entirely from polarization of vacuum fluctuations in the Grid. 
\begin{equation}\label{eqn:9}
\epsilon_0   \sim    \sum_j \frac {N_j (q_j^2/\mu_j)}{ (\omega_0^j)^2 - \omega^2 }  \, .
\end{equation}
For \eqref{eqn:9} to become a defining equation for $\epsilon_0$, the right-hand-side of the above formula must be rewritten so that it describes the interaction of photons with vacuum fluctuations in the Grid that oscillate  instead of oscillators consisting of ordinary matter.  When oscillating vacuum fluctuations in the Grid disappear, they can't leave energy behind for any significant time because the principle of conservation of energy would be violated beyond that allowed by the uncertainty principle. 

The final term $\omega^2/\tau^2$ in the denominator of \eqref{eqn:6a}  occurs because of damping.   While neglecting damping is an approximation for physical particles that oscillate in a dielectric, it is exactly true for  oscillating vacuum fluctuations in the Grid: from  \eqref{eqn:2} it follows that damping arises because (a) oscillators radiate energy  and (b) oscillators lose energy in collisions with other oscillators. But oscillating vacuum fluctuations in the Grid can do neither. If they did, after they vanished they would leave behind  energy, violating the principle of conservation of energy. But this implies that the term multiplying $1/\tau$ in \eqref{eqn:4} vanishes so  $\frac{{\rm d}p_j(t)}{{\rm d}t}=0$ in the Grid.  Consequently, $p_j(t)$ is a constant, implying  $\frac{{\rm d}^2p_j(t)}{{\rm d}t^2}=0$. In the derivation of \eqref{eqn:8}, taking the second derivative of $p_j(t)$  yields the term in the denominator proportional to $\omega^2$. Since the second derivative is zero, the term $\omega^2$ does not appear. Thus  for \eqref{eqn:9} to  describe  the permittivity of the Grid, $\omega^2$ must be set to zero:
\begin{equation}\label{eqn:10}
\epsilon_0 =  \sum_j \frac{(N_j q_j^2/\mu_j)}{(\omega_0^j)^2}\,.
\end{equation}
The above discussion is for a classical electric field. For a quantum field, a photon is absorbed by an oscillating vacuum fluctuation in the Grid. When the  oscillating vacuum fluctuation vanishes into the Grid, a photon is emitted that has the same energy and momentum as the original photon.

The oscillating vacuum fluctuations that are considered here: are the three bound states of a charged lepton and anti-lepton:  positronium, muon-antimuon bound states and tau-antitau bound states.   Initially attention is restricted to positronium.  Conservation of angular momentum requires that positronium be created in the $J=0$ state, which is parapositronium (p-Ps), a singlet spin state  that must decay into an even number of photons. The Heisenberg uncertainty principle is
\begin{equation}\label{eqn:11}
\Delta E_{\rm p-Ps}\, \Delta t_{\rm p-Ps}=\frac{\hbar} {2} \, .
\end{equation}
Denoting the mass of an electron (or positron) by $m_e$,  $\Delta E_{\rm p-Ps}$ is the energy $2m_ec^2$ for the production of  a  parapositronium vacuum fluctuation \footnote{The binding energy of parapositronium, which is small in comparison with $2m_ec^2$, is being neglected.}. Then \eqref{eqn:11} yields the average time $\Delta t_{\rm p-Ps}$ that a parapositronium vacuum fluctuation exists,
\begin{equation}\label{eqn:12}
\Delta t_{\rm p-Ps}= \frac{\hbar}{4m_ec^2} \, .
\end{equation}
During the time $\Delta t_{\rm p-Ps}$, a beam of light  travels a distance $L_{\rm p-Ps}$ given by
\begin{equation}\label{eqn:13}
L_{\rm p-Ps}= c \,\Delta t_{\rm p-Ps}= \frac{\hbar}{4m_ec} \, .
\end{equation}

Here the major new physics, which follows from dimensional analysis and uses standard, cubic wave packets, is the ansatz that  the number of  parapositronium vacuum fluctuations per unit volume is  $1/L_{\rm p-Ps}^3(=1.11\times 10^{39}/{\rm m^3}$), a result that can immediately be generalized to other vacuum fluctuations in the Grid. To minimize the violation of conservation of energy, when a bound particle-antiparticle vacuum fluctuation  is created from the Grid, the pair  appears on mass shell in its lowest energy state. Thus parapositronium, which is positronium's ground state $(n=1)$ for which $J=0$ \cite{Hughes:57}, appears in the Grid as a physically real, time-limited object.  

During the time $\Delta t_{\rm p-Ps}$ that a parapositronium vacuum fluctuation exists, light travels less than one-thousandth of the Bohr radius of parapositronium.  Consequently, a  parapositronium  vacuum fluctuation survives such a short time that when it interacts with a photon, the parapositronium vacuum fluctuation would be expected to vanish back into the Grid before it could be elevated to an excited state.

As in ordinary matter, the decay of a  parapositronium vacuum fluctuation is dominated by its interaction with the quantum electromagnetic field. Therefore, to calculate the probability that a photon interacts with a parapositronium vacuum fluctuation that annihilates electromagnetically, the electromagnetic  decay  rate $\Gamma$ is calculated using the mechanism for the annihilation of ordinary matter, \footnote{Details of this calculation will be given elsewhere.}
\begin{equation}\label{eqn:14} 
\Gamma = \frac{\alpha^5 m_e c^2}{ \hbar} \, ,
\end{equation}
which is twice the decay rate of parapositronium into two photons \cite{Wheeler:46,Jauch:55}.

At equilibrium the average rate for which a parapositronium vacuum fluctuation  absorbs a photon equals the average rate for which  parapositronium vacuum fluctuation  annihilates and emits a photon. As a consequence, the average probability that  a parapositronium vacuum fluctuation absorbs a photon is $\Gamma\, \Delta t_{\rm p-Ps}$.  For a parapositronium vacuum fluctuation the quantity $N_j$ in \eqref{eqn:10}, denoted  $N_{\rm p-Ps}$,  is the number density of  parapositronium  vacuum fluctuations multiplied by the average probability that parapositronium vacuum fluctuation will absorb an incoming photon: 
\begin{equation}\label{eqn:15}
N_{\rm p-Ps} = \frac{1}{L_{\rm p-Ps}^3}\times \Gamma\, \Delta t_{\rm p-Ps}=  \frac{\alpha^5}{4}{\left ({\frac{4 m_e c}{\hbar}} \right )}^3 \,.
\end{equation}

Since positronium is a bound state of an electron and positron, the reduced mass $\mu_i$ in \eqref{eqn:2} is  $\mu_e = m_e/2$.  The  non-relativistic, ground-state energy level for positronium is obtained from the $n = 1$ energy level of hydrogen by replacing the reduced mass of hydrogen with the reduced mass $m_e/2$:
\begin{equation}\label{eqn:16}\
E_e = -\frac{ (m_e/2)e^4 }{2(4\pi \epsilon_0)^2 \hbar^2} =-\frac{m_e\alpha^2c^2}{4} \,.
\end{equation}
The above formula is used  \cite{Holstein:92} to calculate the natural angular frequency of positronium in its ground state: $\omega_0^j=\omega_1^e=-E_e/\hbar$:  
\begin{equation}\label{eqn:17} 
\frac{1}{(\omega_1^e)^2}  =\left(\frac{4\hbar}{m_e\alpha^2c^2}\right)^2\,.
\end{equation}
Eq. \eqref{eqn:10}  then takes the form
\begin{equation}\label{eqn:18}
\epsilon_0 =  \sum_j  \frac{ 8^3\alpha e^2}{\hbar c} \, .
\end{equation}
Note that the mass of the electron has cancelled from the expression for $\epsilon_0$, implying that muon-antimuon vacuum fluctuations and tau-antitau vacuum fluctuations each contribute the same amount to the value of  $\epsilon_0$ as  a parapositronium vacuum fluctuation. Thus,
\begin{equation}\label{eqn:19}
\epsilon_0 = 3\,  \frac{ 8^3\alpha e^2}{\hbar c}\, .
\end{equation}
Multiplying both sides of \eqref{eqn:19} by $(4 \pi \hbar c)/e^2$  and using \eqref{eqn:1} yields the desired result:
\begin{equation}\label{eqn:20}
\frac{1}{\alpha} \cong 8^2\sqrt{3\pi/2}\cong 138.93\dots \, .
\end{equation}
The experimental value is $1/\alpha=137.036\dots$. Using the second expression for $\alpha$ in \eqref{eqn:1}, substituting the expression into \eqref{eqn:19}, and solving for $\epsilon_0$,  
\begin{subequations}\label{eqn:21}
\begin{equation}\label{eqn:21a}
\epsilon_0 \cong 4^2\sqrt{6\pi}\,\frac{e^2}{hc} \cong 8.98\times 10^{-12}\frac{{\rm C^2}}{{\rm Nm^2}} \, .
\end{equation}
Alternatively, selecting the third expression for $\alpha$ in \eqref{eqn:1} that includes the defined quantity $\mu_0$, substituting the expression into \eqref{eqn:19}, and solving for $\epsilon_0$,  
\begin{equation}\label{eqn:21b}
\epsilon_0 \cong 3(8^3) \pi\,\frac{\mu_0e^4}{h^2} \cong 9.10\times 10^{-12}\frac{{\rm C^2}}{{\rm Nm^2}} \, .
\end{equation}
\end{subequations}
Because the expression for $\epsilon_0$ in \eqref{eqn:19} is not exact,  slightly different values for $\epsilon_0$ are obtained when expressed in terms of  $\mu_0, \,e$, and $h$  instead of $c, \,e,$ and $h$.  The experimental  value is $\epsilon_0=8.85\times 10^{-12} {\rm C^2}/{\rm Nm^2}$. Using  $c=1/\sqrt{\mu_0\epsilon_0}$ and the value for $\epsilon_0$ in \eqref{eqn:21a}, $c=2.98\times 10^8$m/s.

Vacuum fluctuations consisting of quark-antiquark pairs contribute little to the value of $\alpha$: for the heavy quarks $Q=c, b,\, {\rm or}\, t$, there are no $Q{\bar Q}$ states that decay directly into two photons.  While there are $Q{\bar Q}$ states that decay into a lepton-antilepton pair that, in turn, decays into two photons, the contributions to $\epsilon_0$ from these decays are suppressed because of the intermediate lepton-antilepton pair. For the light quarks $q=u, d,\, {\rm or}\, s$, the $\pi^0, \eta$, and $\eta^\prime$  are $J=0$ combinations of $q{\bar q}$ bound states that decay into two photons.  If the the $\pi^0, \eta$, and $\eta^\prime$ resonate, the frequencies would be many orders of magnitude larger than those of bound, lepton-antilepton pairs, suppressing their contributions to $1/\alpha$. On the other hand, in comparison to their masses, the decay rates of these mesons are orders of magnitude larger than those of bound, lepton-antilepton pairs. The net result is that the light quarks' contributions  also would not significantly affect the value of $\alpha$ to the accuracy obtained here.

The fact that the above value for $1/\alpha$ is  about 1.4\%  too large might be explained  if some lepton-antilepton vacuum fluctuations convert into another form  that contributes less to the the right-hand side of \eqref{eqn:10}.  Specifically, if some  parapositronium vacuum fluctuations  combine to form vacuum fluctuations of diparapositronium molecules that have an electric quadrupole moment but not a dipole moment, then the  right-hand side of \eqref{eqn:10} and thus $1/\alpha$ would decrease. Also, higher-order corrections in $\alpha$ could be significant.   

Note in particular that diparapositronium molecules appearing as vacuum fluctuations respond to (quadrupole) gravitational radiation, inextricably linking electromagnetism and gravitational radiation. This subject is currently under investigation.

In the early universe when the temperature was sufficiently high that it was difficult for lepton-antilepton  vacuum fluctuations to form, the number density of  lepton-antilepton vacuum fluctuations in the Grid would have been much less than today. From \eqref{eqn:10} it then follows that $\epsilon_0$ would also have been much smaller.  Since $c=1/\sqrt{\epsilon_0 \mu_0}$,  the decrease in the value of $\epsilon_0$ would tend to make the speed of light in the early universe much larger. 

\bibliography{Alpha2}
\end{document}